\documentclass[aps,pre,a4paper,twocolumn,groupedaddress,showpacs,showkeys,preprintnumbers,floatfix,dvips,superscriptaddress]{revtex4-1}

\usepackage[utf8]{inputenc} \usepackage[T1]{fontenc}
\usepackage{psfrag} 
\usepackage{times} \usepackage{graphicx} \usepackage{color}
\usepackage{amsmath} \usepackage{amscd} \usepackage{amssymb} \usepackage{amsthm}
\usepackage{ragged2e} \usepackage{subfigure} \usepackage{caption}
\usepackage{nicefrac}
\usepackage{captcont}
\usepackage{verbatim}

\DeclareCaptionJustification{reallyjustified}{\justifying}
\begin{document}

\title{Testing the validity of the Kirkwood approximation using an extended Sznajd model.}
\author{Andr\'e M. Timpanaro}
\email[]{timpa@if.usp.br}
\affiliation{Instituto de F\'{i}sica, Universidade de S\~{a}o Paulo Caixa Postal 66318, 05314-970 - S\~{a}o Paulo - S\~{a}o Paulo - Brazil}
\affiliation{CEVIPOF - SciencesPo and CNRS -  Paris, 98 rue de l'Universit\'e, 75007 - France}
\author{Serge Galam}
\email[]{serge.galam@sciencespo.fr}
\affiliation{CEVIPOF - SciencesPo and CNRS -  Paris, 98 rue de l'Universit\'e, 75007 - France}
\date{\today}

\pacs{}

\begin{abstract}

We revisit the deduction of the exit probability of the one dimensional Sznajd model through the Kirkwood approximation [F. Slanina \emph{et al.}, Europhys. Lett. {\bf 82}, 18006 (2008)]. This approximation is peculiar in that in spite of the agreement with simulation results [F. Slanina \emph{et al.}, Europhys. Lett. {\bf 82}, 18006 (2008); R. Lambiotte and S. Redner, Europhys. Lett. {\bf 82}, 18007 (2008); A. M. Timpanaro and C. P. C. Prado, {\bf 89}, 052808 (2014)] the hypothesis about the correlation lenghts behind it are inconsistent and fixing these inconsistencies leads to the same results as a simple mean field. We use an extended version of the Sznajd model to test the Kirkwood approximation in a wider context. This model includes the voter, Sznajd and ``United we Stand, Divided we Fall'' (USDF) models [R. A. Holley and T. M. Liggett, Ann. Prob. {\bf 3}, 643 (1975); K. Sznajd-Weron and J. Sznajd, Int. Journ. Mod. Phys. C {\bf 11}, 1157 (2000)] as different parameter combinations, meaning that some analytical results from these models can be used to evaluate the performance of the Kirkwood approximation. We also compare the predicted exit probability with simulation results for networks with $10^3$ sites. The results show clearly the regions in parameter space where the approximation gives accurate predictions, as well as where it starts failing, leading to a better understanding of its reliability.


\end{abstract}

\maketitle

\section{Introduction}

In this work we investigate the reliability of the Kirkwood approximation, as used in \cite{q-voter-Slanina} as a tool to study agent models defined in one dimensional lattices. This approximation is a type of mean field approach that keeps pair correlations but makes some unusual hypothesis about how the correlations decay. It was used in \cite{q-voter-Slanina} to study the exit probability of the Sznajd opinion propagation model (the probability of reaching one of the two possible absorbing states as a function of the initial conditions) and a controversy about wether these results are indeed correct emerged in a series of papers \cite{q-voter-Slanina, q-voter-Sznajd, q-voter-Lambiotte, q-voter-Galam, q-voter-Timpanaro, Timpanaro-Galam-q-voter}. Simulation results later stablished that the predicted exit probability is impressively acurate (it has been confirmed for simulations in networks with up to $3.16\times 10^7$ sites \cite{q-voter-Timpanaro}) and the same expression has since been obtained by a non mean field treatment that makes some hypothesis about how large groups of sites having the same opinion interact with each other \cite{Timpanaro-Galam-q-voter}. Also, the fact that a mean field treatment manages to give an acurate answer in an one dimensional problem raised some speculations about wether there is something special about the way correlations behave in this model or if it is some kind of coincidence \cite{castellano-kirkwood-bashing}.

Motivated by this, we study the Kirkwood approximation in a wider context. We will focus on using this approximation to study the exit probability of a generalization of the Sznajd model that includes also the voter and ``United we Stand, Divided we Fall'' (USDF) models \cite{kondrat-two-components, votante-def, sznajd-def}, the reasoning being that this would create a parameter space where we can study the model. As we know that the approximation is acurate for some parameter values and expect this to be true for small perturbations of these parameters, this would lead to regions where the approximation works and regions where it fails, improving the understanding of what is behind it.

We will first define precisely the model used and make calculations for the exit probability (section \ref{sec:Kondrat-def}) and later compare the obtained expression with simulation results (section \ref{sec:sim-res}).

\subsection{The approximation in the Sznajd model}

The Sznajd model is an opinion propagation model, derived from the Ising model. In the situations that we will consider, the model is defined in an one dimensional chain with $N$ sites and periodic boundary conditions. Each of its sites represents a person that can have one of two possible spin states, representing opinions (denoted $+$ and $-$). The time evolution is defined as follows:
\begin{itemize}
\item At each time step, choose at random a pair of neighbouring sites $i$ and $i+1$.
\item If their opinions $\sigma_i$ and $\sigma_{i+1}$ are equal then we choose one of the neighbours of the pair, $i-1$ or $i+2$ and change its opinion to the pair opinion $\sigma_i$.
\item Otherwise, if $\sigma_i \neq \sigma_{i+1}$, nothing happens and we move on to the next time step.
\end{itemize}

The Kirkwood approximation can be summed up as
\begin{itemize}
\item Break correlations by pairing neighbouring spins and considering spins in a pair independent from spins in other pairs:
\[
\langle\sigma_1\sigma_2\sigma_3\rangle \simeq \langle\sigma_1\sigma_2\rangle\langle\sigma_3\rangle
\]
\item However, when we have only 2 spins we consider then correlated as if they were nearest neighbours:
\[
\langle\sigma_1\sigma_4\rangle \simeq \langle\sigma_1\sigma_2\rangle.
\]
\end{itemize}
This unusual aspect of considering correlation lenghts to be short and long at the same time, depending on which correlation is being considered casts doubt about wheter the approximation scheme even makes sense. However as we will show, assuming a more coherent way of reducing the correlations (namely $\langle\sigma_i\sigma_{i+n}\rangle \simeq \langle\sigma_i\rangle\langle\sigma_{i+n}\rangle$) reduces the results to the same as a simple mean field approximation (that gives a simpler, but inacurate, exit probability).

We start by considering the master equation of the model. Since the only changes that are possible in each time step are single spin flips, then

\begin{equation}
\dot{P}(\sigma) = \sum_{i}\left(W_i(\sigma^i) P(\sigma^i) - W_i(\sigma) P(\sigma)\right),
\label{eq:markov-base}
\end{equation}
where $\sigma$ denotes the state of the chain, $\sigma^i$ is the state $\sigma$ flipping the site $i$, $\nicefrac{W_i(\sigma)}{N}$ is the probability of going from state $\sigma$ to state $\sigma^i$ in one time step and the sum over $i$ runs through all the sites in the chain. Consider then a function $\varphi(\sigma)$. Its ensemble average is given by

\[
\langle\varphi\rangle = \sum_{\sigma} \varphi(\sigma)P(\sigma),\quad\mbox{so}\quad\frac{\mathrm{d}}{\mathrm{dt}}\langle\varphi\rangle = \sum_{\sigma} \varphi(\sigma)\dot{P}(\sigma).
\]
Substituting the master equation and reordering the sums leads to

\[\frac{\mathrm{d}}{\mathrm{dt}}\langle\varphi\rangle = \sum_{\sigma}P(\sigma)\sum_{i}W_i(\sigma)(\varphi(\sigma^i) - \varphi(\sigma)) = \]
\begin{equation}
= \left\langle\sum_{i}W_i(\sigma)(\varphi(\sigma^i) - \varphi(\sigma))\right\rangle.
\label{eq:aux-a}
\end{equation}
We will be particularly interested in studying the average of spin products

\begin{equation}
\varphi(\sigma) = \prod_{k=1}^n \sigma_{a_k}.
\label{eq:aux-b}
\end{equation}
As $\sigma_i^j = \sigma_i(1-2\delta_{i,j})$, substituting eq \ref{eq:aux-b} in \ref{eq:aux-a} leads to

\begin{equation}
\frac{\mathrm{d}}{\mathrm{dt}}\left\langle\prod_{k=1}^n \sigma_{a_k}\right\rangle = -2\sum_{k=1}^n\left\langle W_{a_k}(\sigma)\prod_{q=1}^n \sigma_{a_q}\right\rangle.
\label{eq:markov-prod}
\end{equation}
So to obtain the time evolution equations we must find $W_i$ and use equation \ref{eq:markov-prod} to obtain an hierarchy of equations that will be truncated by the Kirkwood approximation. For the Sznajd model we have

\[W_i(\sigma) = \frac{1}{8}
(2 - \sigma_i(\sigma_{i+1} + \sigma_{i-1}) - \sigma_i(\sigma_{i+2} + \sigma_{i-2}) +
\]
\begin{equation}
+\sigma_{i+1}\sigma_{i+2} + \sigma_{i-1}\sigma_{i-2}).
\label{eq:W-sznajd}
\end{equation}
Defining the following spin averages

\[
C_1 = \langle \sigma_i \rangle ,\quad C_2(k) = \langle \sigma_i \sigma_{i+k} \rangle ,\quad 
\]
\begin{equation}
C_3 = \langle \sigma_i\sigma_{i+1}\sigma_{i+2} \rangle \quad \mathrm{and} \quad C_4 = \langle \sigma_i\sigma_{i+1}\sigma_{i+2}\sigma_{i+3} \rangle
\end{equation}
it follows from equations \ref{eq:markov-prod} and \ref{eq:W-sznajd} that

\begin{equation}
\left\{
\begin{array}{l}
\dot{C}_1 = C_1 - C_3 \\
\dot{C}_2(1) = 1 - C_2(1) + C_2(3) - C_4
\end{array}
\right.
\label{eq:markov-sznajd}
\end{equation}
Applying the Kirkwood approximation leads us to the following

\begin{equation}
\left\{
\begin{array}{l}
C_3 \simeq C_2(1)C_1 \\
C_4 \simeq C_2(1)^2 \\
C_2(3) \simeq C_2(1)
\end{array}
\right.
\label{eq:kirkwood}
\end{equation}
Since $C_2(1)$ is the only $C_2(k)$ that remains we will shorthand it to $C_2$. It follows that

\begin{equation}
\left\{
\begin{array}{l}
\dot{C}_1 = C_1(1 - C_2) \\
\dot{C}_2 = 1 - C_2^2
\end{array}
\right.
\label{eq:kirk_sznajd}
\end{equation}
We want to calculate the exit probability, which is the probability of reaching the absorbing state where all spins equal $+$ as a function of the initial proportion $\rho$ of sites $+$ (assuming also that the initial condition is completely uncorrelated). Since $C_1 = 1$ in the all $+$ state and $C_1 = -1$ in the all $-$ state (which are the only relevant absorbing states), we can obtain the exit probability by calculating the limit

\[
\lim_{t\rightarrow \infty}\frac{1 + C_1(t)}{2}
\]
and using $C_1(t=0) = 2\rho - 1$ and $C_2(t=0) = C_1(t=0)^2 = (2\rho-1)^2$ as initial conditions. Doing this leads us to the following expression

\begin{equation}
E(\rho) = \frac{\rho^2}{\rho^2 + (1-\rho)^2},
\label{eq:EP-q2}
\end{equation}
which matches the simulation results fairly well. However, as we pointed already, the Kirkwood approximation is inconsistent in the way it reduces correlations and this inconsistency is crucial for the result in eq \ref{eq:EP-q2}. 
Obviously, if we assume that the correlations have longer ranges than just the first neighbour then we would need to treat $C_2(3)$, $C_3$ and $C_4$ as separate variables and derive more equations through relation \ref{eq:markov-prod}, involving more complicated spin averages and yielding a much more complex treatment. On the other hand if we reduce $C_2(k)$ in a way that is coherent with how the other correlations were treated then one should use $C_2(k) \simeq C_1^2$ when $k>1$. If we substitute $C_2(3) = C_1^2$ in eqs \ref{eq:markov-sznajd} we get this system of equations instead

\begin{equation}
\left\{
\begin{array}{l}
\dot{C}_1 = C_1(1 - C_2) \\
\dot{C}_2 = 1 + C_1^2 - C_2 - C_2^2
\end{array}
\right.
\label{eq:coherent_sznajd}
\end{equation}
and integrating these new equations leads to an exit probability given by a step function

\[
E(\rho) = \Theta \left(\rho - \frac{1}{2}\right),
\]
which is the same result as using a simple mean field ($\dot{C}_1 = C_1 - C_1^3$), as well as applying the Galam Unified Frame to this problem \cite{Galam-2005}.










\section{The extension of the model}
\label{sec:Kondrat-def}

The extension of the Sznajd model that we will use was proposed by Kondrat in \cite{kondrat-two-components} and is defined by the following evolution rules

\begin{itemize}
\item We choose a site $i$ at random and take either the two neighbours to the left or the two to the right (with $50\%$ probability each) to form a triplet. We will denote the nearest neighbour $j$ and the second nearest $k$.
\item We flip site $i$ ($\sigma_i \rightarrow -\sigma_i$) with probability $p(\sigma_i, \sigma_j, \sigma_k)$ and move on to the next time step.
\end{itemize}

It follows that the model is completely specified by the $p$ function. Since we have 2 possible spin states for each site there are 8 possible triplets $(\sigma_i, \sigma_j, \sigma_k)$. However, we are only interested in the situation where the $+$ and $-$ spins behave in a symmetric way, so we impose the constraint

\[
p(\sigma_i, \sigma_j, \sigma_k) = p(-\sigma_i, -\sigma_j, -\sigma_k),
\]
leaving us with 4 parameters to characterize $p$. These parameters can be given an interpretation in terms of the type of opinion change they represent:

\[
\begin{array}{l}
p_1 = p(+, +, +), \quad\mbox{contrarian behaviour}\\
p_2 = p(-, +, +), \quad\mbox{conformity behaviour}\\
p_3 = p(+, -, +), \quad\mbox{conformity behaviour}\\
p_4 = p(-, -, +), \quad\mbox{disagreement propagation}.
\end{array}
\]
Moreover, 3 known models correspond to specific parameter choices:

\[
\begin{array}{l}
p_2 = 1, p_1 = p_3 = p_4 = 0 \quad\mbox{leads to the Sznajd model}\\
p_2 = p_3 = 1, p_1 = p_4 = 0 \quad\mbox{leads to the voter model}\\
p_2 = p_4 = 1, p_1 = p_3 = 0 \quad\mbox{leads to the USDF model}.
\end{array}
\]
We want to find a system of equations similar to eqs \ref{eq:kirk_sznajd}, meaning we must obtain an expression for $W_i(\sigma)$. We first note that for these spin variables $\delta_{a,b,c} = \nicefrac{(1+ab+ac+bc)}{4} \equiv \Delta(a,b,c)$, meaning that
\[
\xi_i(a,b,c) = \frac{\Delta(a\sigma_i, b\sigma_{i-1}, c\sigma_{i-2}) + \Delta(a\sigma_i, b\sigma_{i+1}, c\sigma_{i+2})}{2}
\]
can be used to match the different spin patterns $(\pm,\pm,\pm)$, $(\mp,\pm,\pm)$, $(\pm,\mp,\pm)$ and $(\mp,\mp,\pm)$, following that
\begin{equation}
W_i(\sigma) = \sum_{a,b = \pm 1} p(a,b,+)\xi_i(a,b,+).
\label{eq:w-generico}
\end{equation}
Expanding eq \ref{eq:w-generico} leads to

\[
W_i(\sigma) = 2\alpha + \beta\sigma_i(\sigma_{i+1} + \sigma_{i-1}) + \gamma\sigma_i(\sigma_{i+2} + \sigma_{i-2}) + 
\]
\begin{equation}
+\delta(\sigma_{i+1}\sigma_{i+2} + \sigma_{i-1}\sigma_{i-2})
\end{equation}
where $\alpha, \beta, \gamma$ and $\delta$ are the following parameter combinations
\[
\alpha \equiv \frac{p_1+p_2+p_3+p_4}{8}
\]\[
\beta  \equiv \frac{p_1-p_2-p_3+p_4}{8}
\]\[
\gamma \equiv \frac{p_1-p_2+p_3-p_4}{8}
\]\[
\delta \equiv \frac{p_1+p_2-p_3-p_4}{8}.
\]
Repeating what we did for the Sznajd model leads to

\begin{equation}
\left\{
\begin{array}{lcl}
\dot{C}_1 & = & -4((\alpha + \beta + \gamma)C_1 + \delta C_3) \\
\dot{C}_2(1) & = & -4(\beta + (2\alpha + \gamma)C_2(1) + (\beta + \delta)C_2(2) \\
& &  + \gamma C_2(3) + \delta C_4)
\end{array}
\right.
\label{eq:markov-kondrat}
\end{equation}
and after applying the approximation

\begin{equation}
\left\{
\begin{array}{ll}
\dot{C}_1  & = -4C_1(\alpha + \beta + \gamma + \delta C_2) \\
\dot{C}_2 & = -4(\beta + (2\alpha + \beta + 2\gamma + \delta)C_2 + \delta C_2^2)
\end{array}
\right.
\label{eq:markov-kondrat-1}
\end{equation}
In order for the model to have an exit probability we need it to order, meaning that $C_2$ must go to 1 as time goes by. However, substituting $C_2 = 1$ in the equation for its derivative yields $\dot{C}_2 = -8(\alpha + \beta + \gamma + \delta) = -4p_1$. Hence, we must restrict ourselves to models such that $p_1 = 0$ (which makes sense, because the ferromagnetic states are not absorbing otherwise). As this means that $\alpha + \beta + \gamma + \delta = 0$ the equations can be further simplified:

\begin{equation}
\left\{
\begin{array}{ll}
\dot{C}_1  & = 4\delta C_1(1 - C_2) \\
\dot{C}_2 & = -4(\beta - (\beta + \delta)C_2 + \delta C_2^2)
\end{array}
\right.
\label{eq:markov-kondrat-2}
\end{equation}
We also want that the solution $C_2 = 1$ be attractive, which means $2\delta - (\beta + \delta) > 0 \Leftrightarrow \delta > \beta \Leftrightarrow p_2 > p_4$. With these two conditions in place the integration of the equations \ref{eq:markov-kondrat-2} becomes very similar to what was done in the case of the Sznajd model and we can obtain the following exit probability

\begin{equation}
E(\rho) = \frac{\rho(1-\rho)p_3 + \rho^2(p_2-p_4)}{2\rho(1-\rho)p_3 + (\rho^2 + (1-\rho)^2)(p_2-p_4)}.
\label{eq:EP-kondrat}
\end{equation}

Once again we can try substituting $C_2(2)$ and $C_2(3)$ by $C_1^2$ and the exit probability obtained is the same as in the case of a mean field with no correlations (i.e. integrating $\dot{C}_1 = 4\delta C_1(1 - C_1^2)$), with a step function for $\delta < 0$, a linear exit probability for $\delta = 0$ and the model failing to order if $\delta > 0$.







\section{Simulation results}
\label{sec:sim-res}

We now make simulations of the model to compare the results with the expression \ref{eq:EP-kondrat}. Since we are restricting ourselves to $p_1 = 0$ and $p_2 > p_4$ we have 3 free parameters: $p_2$, $p_3$ and $p_4$. However the parameter space we need to explore has only 2 dimensions since the model with parameters $p'_i = \mu p_i$ differs from the model with parameters $p_i$ only on the time scale where things happen, so that the limit for large times remains the same (this can be seen in equation \ref{eq:EP-kondrat} that is invariant by a rescaling of the parameters). As a consequence, we only need to analyse two different cases: $p_2 = 1$ with $p_3$ and $p_4$ free, and $p_3 = 1$ with $p_2 > p_4$.

\begin{figure}
\includegraphics[width=\columnwidth]{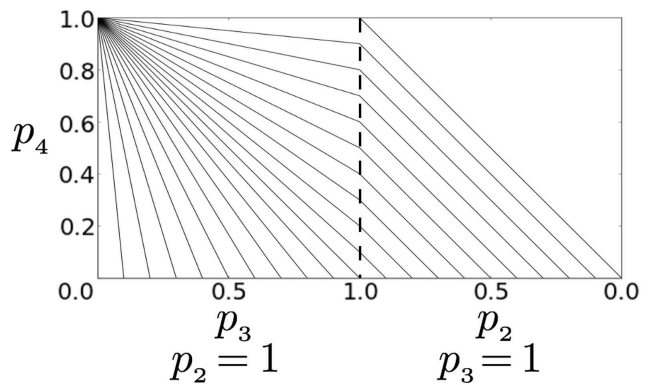}
\caption{Curves along which $\psi = \nicefrac{p_3}{(p_2 - p_4)}$ is a constant in parameter space. Along each curve the functional form of $E(\rho)$ predicted by the Kirkwood approximation is the same. Note the singularity for $p_2 = p_4 = 1$ and $p_3 = 0$ that corresponds to the USDF model.}
\label{fig:iso-EP}
\end{figure}

Before making the simulations we take a closer look at the prediction given by the Kirkwood approximation, eq \ref{eq:EP-kondrat}. Since the expression is invariant by a reescaling of the parameters and only 2 parameter combinations appear ($p_3$ and $p_2-p_4$) we can rewrite it using only one parameter $\psi$ and the dependance of the exit probability with $\rho$ would be the same along lines with constant $\psi$. We choose

\[
\psi = \frac{p_3}{p_2 - p_4},
\]
that varies from 0 when $p_2 = 1$ and $p_3 = 0$ and diverges in the limit $p_2 \rightarrow p_4$, when the Kirkwood approximation becomes problematic. The exit probability in eq \ref{eq:EP-kondrat} becomes

\begin{equation}
E(\rho) = \frac{\psi\rho(1-\rho) + \rho^2}{2\psi\rho(1-\rho) + \rho^2 + (1-\rho)^2}
\label{eq:EP-psi}
\end{equation}
and the curves $\psi = \mathrm{constant}$ can be seen in figure \ref{fig:iso-EP}. Note the singularity that arises for $p_2 = p_4 = 1$ and $p_3 = 0$, corresponding to the USDF model (that has antiferromagnetic states with a non-zero probability of being reached in addition to the ferromagnetic ones, meaning that the exit probability as we defined is not adequate in this case).

We also know that the model with parameters such that $p_2 = p_3 + p_4$ (corresponding to $\psi = 1$) must have a linear exit probability (excluding the USDF case). This happens because the model can be rewritten (up to a time scale) as follows:
\begin{itemize}
\item At each time step choose a site $i$.
\item With probability $\nicefrac{p_3}{p_2}$, $i$ copies the opinion of one of its first neighbours.
\item Otherwise (hence with probability $\nicefrac{p_4}{p_2}$) $i$ copies the opinion of one of its second neighbours.
\end{itemize}
Since both of the possible interactions conserve the magnetization ($N_+ - N_-$) on average, the ensemble average of $\rho$ after very long times (corresponding to the exit probability in these cases) is the same as in the initial condition, implying $E = \rho$. Because of this we didn't made simulations in this case, using the analytical result instead.

\begin{figure}
\includegraphics[width=\columnwidth]{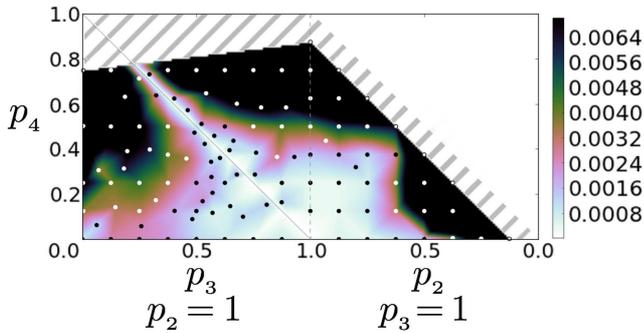}
\caption{The graph shows the estimated absolute error of the Kirkwood approximation for each of the simulated parameter sets. Brighter colors indicate smaller differences, while darker colors indicate larger differences (values larger than $0.007$ are all in black). The marked points indicate the data we have and that was interpolated to yield the graph, while the hashed region corresponds to regions where we had no data. The line $p_2 = p_3 + p_4$ is also marked, where the error was set to 0.}
\label{fig:grafico-duplo-abs}
\end{figure}

For the simulations we used linear chains with periodic conditions and $10^3$ sites. While this is a relatively small network size it is enough to get a rough picture of how the exit probability behaves (the main reason larger network sizes were not used is that the techniques used to speed up the results for the Sznajd model \cite{q-voter-Timpanaro} are not as efficient when $p_4 \neq 0$). For each parameter set we made simulations using $\rho = 0.1, 0.2, 0.3, 0.4$ and $0.45$, calculating the absolute difference between the simulation results for the exit probability and the one predicted with the Kirkwood approximation and to visualize these in two dimensions we made the average of this absolute difference (that works as an estimate of the absolute error). Finaly, depending on the parameters, we made between $10^5$ and $10^6$ simulations (depending mainly on how long the simulations took and how much the measured exit probability deviated from the one predicted by the Kirkwood approximation). The results obtained are summed up in figure \ref{fig:grafico-duplo-abs}.

We can distinguish easily that the Kirkwood approximation starts failing as $p_4\rightarrow p_2$, as well as two distinct regions where the approximation has greater accuracy. One of them is the small region around the Sznajd model ($p_2 = 1, p_3 = p_4 = 0$, corresponding to the lower left corner in the graph of figure \ref{fig:grafico-duplo-abs}) and the other contains the vicinity of the voter model ($p_2 = p_3 = 1, p_4 = 0$) as well as the line $p_2 = p_3 + p_4$ and part of $p_3 = p_2 + p_4$.

\begin{figure}
\includegraphics[width=\columnwidth]{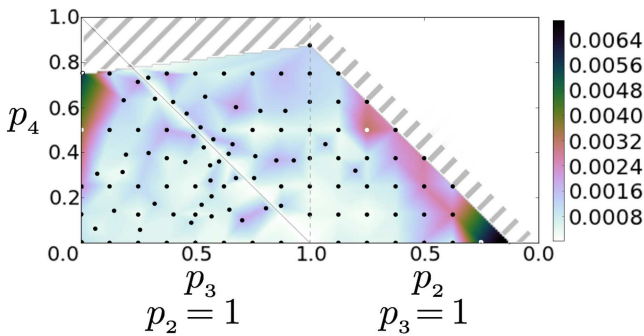}
\caption{The graph shows the estimated absolute error of equation \ref{eq:EP-psi} with a fitted value of $\psi$, for each of the simulated parameter sets. Brighter colors indicate smaller differences, while darker colors indicate larger differences (values larger than $0.007$ are all in black). The marked points indicate the data we have and that was interpolated to yield the graph, while the hashed region corresponds to regions where we had no data. The line $p_2 = p_3 + p_4$ is also marked, where the error was set to 0.}
\label{fig:duplo-psi}
\end{figure}

Nevertheless, equation \ref{eq:EP-psi} alone gives a good description of the exit probability in most of the parameter space if we allow $\psi$ to take values different than the ones predicted by the Kirkwood approximation. We repeated the comparison made in figure \ref{fig:grafico-duplo-abs}, but now using a fitted value of $\psi$ instead of the one predicted by the approximation. This comparison can be found in figure \ref{fig:duplo-psi}, where we can see that most of the parameter space seems to obey the functional form in eq \ref{eq:EP-psi} even though only a part of it matches the exact prediction of the Kirkwood approximation.

\begin{figure}
\includegraphics[width=\columnwidth]{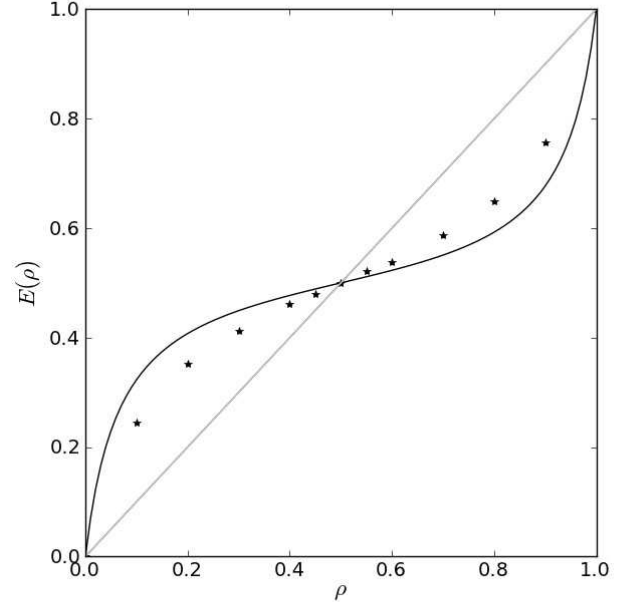}
\caption{Simulation results ($\star$) and the Kirkwood approximation (darker curve) for the parameter choice $p_2 = p_3 = 1$, $p_4 = 0.875$. The curve $E=\rho$ is drawn for reference (lighter line). Statistical errors are too small to see.}
\label{fig:exotica1}
\end{figure}

\begin{figure}
\includegraphics[width=\columnwidth]{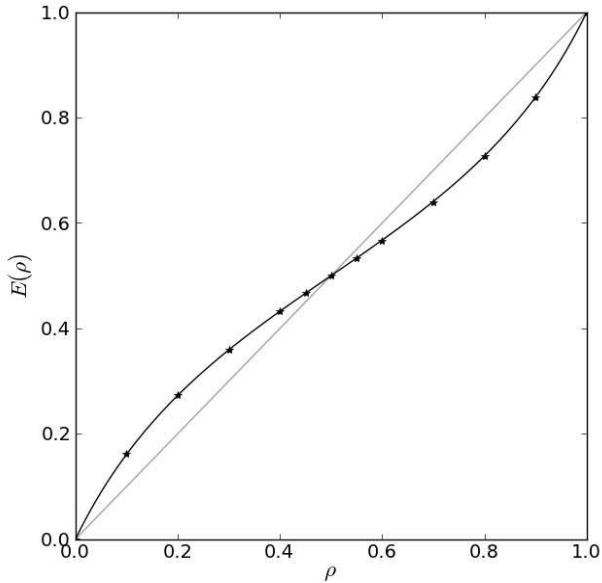}
\caption{Simulation results ($\star$) and the Kirkwood approximation (darker curve) for the parameter choice $p_2 = 0.75$, $p_3 = 1$, $p_4 = 0.25$. The curve $E=\rho$ is drawn for reference (lighter line). Statistical errors are too small to see.}
\label{fig:exotica2}
\end{figure}

Finally we call to attention the peculiar aspect of the exit probability when $\psi > 1$ (which corresponds to more than half of the parameter space), where the minority proportion always increases, but the model still orders. It is important to note that this means that an uncorrelated initial condition will see an increase of the minority opinion, while this ceases to be true once correlations form (allowing the system to order). This must be an effect of simulating the model in one dimension and shouldn't happen in networks offering more possibilities for the opinions to mix. Figures \ref{fig:exotica1} and \ref{fig:exotica2} show the Kirkwood approximation and simulation results for two parameter choices (one where the approximation is acurate and other where it is not).

\section{Conclusion}

In this work we approached the controversy about the exit probability of the Sznajd model through a different angle, aiming at the validity of the Kirkwood approximation itself. Our simulations show some clear regimes where the simulation does not yield accurate results, but nevertheless it seems to give correct qualitative results (like the region in the parameter space where the system orders and the general functional form of the exit probability), as would be expected from a mean field treatment. We have also shown that the way correlations are approximated appears to be inconsistent and that if we make instead consistent choices, the conclusions about the exit probability and the parameters for which the system orders match the ones obtained by a simple mean field approximation. On the other hand the approximation also gives remarkably acurate predictions in some parameter regions. This acuracy was expected for the line $p_2 = p_3 + p_4$ as we pointed out, because the magnetization is conserved in average so the correlations are not all that important. The large region in the center of figure \ref{fig:grafico-duplo-abs} however was completely unexpected and may give a hint about when the inconsistencies in the treatment of the correlations turn out to be really relevant. In particular, part of the region where $p_2 < p_3 + p_4$ has a peculiar shape for the exit probability curve that has been confirmed by the simulations (see figure \ref{fig:exotica2}) and that would imply that minority opinions become represented disproportionately more than majority opinions. Situations like this should not intuitively happen unless the model failed to order. In our opinion this is an effect of simulating the model in one dimension, which severely curtails how much the opinions can mix. What is also remarkable is that the Sznajd model seems to be in a small region of acuracy,which could indicate that there really is something special about the way correlations behave in it.

\bibliographystyle{plain}
\bibliography{andre}

\end{document}